\documentclass[10pt,aps,twocolumn,showpacs,preprintnumbers,amsmath,amssymb,prb,superscriptaddress,floatfix]{revtex4}

\usepackage{graphicx}
\usepackage{amsmath,amsfonts,amsthm}
\usepackage{bbm}
\usepackage{amssymb}

\newcommand{\expct}[1]{\left\langle #1 \right\rangle}
\newcommand{\expcts}[1]{\langle #1 \rangle}
\newcommand{\cumu}[1]{\langle\!\langle #1 \rangle\!\rangle}
\newcommand{\cumus}[1]{\langle\!\langle #1 \rangle\!\rangle}
\newcommand{\ket}[1]{\left| #1 \right\rangle}

\newcommand{\bra}[1]{\left\langle #1 \right|}

\newcommand{\hn}{\hat{n}}

\DeclareMathOperator{\sgn}{sgn}
\DeclareMathOperator{\Tr}{Tr}
\DeclareMathOperator{\hc}{h.c.}

\newcommand{\hk}{K}
\newcommand{\hj}{J}

\begin{document}

\title{Detection of qubit-oscillator entanglement in nanoelectromechanical systems}
\author{Thomas~L.~Schmidt}
\email{t.schmidt@yale.edu}
\affiliation{Department of Physics, Yale University, 217 Prospect Street, New Haven, Connecticut 06520, USA}
\affiliation{Department of Physics, University of Basel, CH-4056
  Basel, Switzerland}
\author{Kjetil B\o{}rkje}
\affiliation{Department of Physics, Yale University, 217 Prospect Street, New Haven, Connecticut 06520, USA}
\author{Christoph Bruder}
\affiliation{Department of Physics, University of Basel, CH-4056
  Basel, Switzerland}
\author{Bj\"orn Trauzettel}
\affiliation{Institute for Theoretical Physics and Astrophysics,
University of W\"urzburg, D-97074 W\"urzburg, Germany}

\date{\today}

\begin{abstract}
Experiments over the past years have demonstrated that it is possible to bring nanomechanical resonators and superconducting qubits close to the quantum regime and to measure their properties with an accuracy close to the Heisenberg uncertainty limit. Therefore, it is just a question of time before we will routinely see true quantum effects in nanomechanical systems. One of the hallmarks of quantum mechanics is the existence of entangled states. We propose a realistic scenario making it possible to detect entanglement of a mechanical resonator and a qubit in a nanoelectromechanical setup. The detection scheme involves only standard current and noise measurements of an atomic point contact coupled to an oscillator and a qubit. This setup could allow for the first observation of entanglement between a continuous and a discrete quantum system in the solid state.
\end{abstract}


\maketitle

In recent years, nanoelectromechanical systems (NEMS) have become a strong focus of research in theoretical and experimental physics.\cite{schwab05} One of the practical reasons for this development is the prospective use of NEMS to design devices which allow the measurement of position, force and mass\cite{lahaye04,mamin01,yang06} with unprecedented accuracies. From a more fundamental point of view, NEMS operate at the boundary between the classical and the quantum world and recent works have suggested that NEMS will soon allow the observation of quantum mechanical states in mesoscopic mechanical systems.\cite{lahaye09}

The observation of quantum states of matter in such systems generally requires ultralow temperature and low dissipation. Bringing a nanomechanical oscillator of frequency $\Omega$ near its ground state means reaching a temperature $T \ll \hbar \Omega/k_B$. Various schemes to cool an oscillator to its ground state have been proposed,\cite{courty01,wilson-rae07,marquardt07} and experiments on nanomechanical systems are now approaching this limit.\cite{naik06,rocheleau09} Moreover, high quality factors have been achieved which lead to relaxation and decoherence times long enough for the measurement of quantum states.\cite{huettel09,steele09}

One of the most rewarding endeavors involves the creation and detection of nonclassical correlations (entanglement) between the nanomechanical oscillator and another quantum system. The easiest option would be to entangle the oscillator with a mesoscopic system whose properties are well understood and in which quantum effects can routinely be observed: a superconducting qubit.\cite{bouchiat98,nakamura99} It has been demonstrated that these devices have decoherence times which can exceed oscillation periods of nanomechanical resonators by several orders of magnitude.\cite{wallraff05} Various theoretical proposals have been made on how entanglement between an oscillator and a qubit can be created\cite{armour02,tian05} and such systems have been successfully coupled in experiments.\cite{lahaye09,hofheinz09}

\begin{figure}[t]
  \centering
  \includegraphics[width = 0.47\textwidth]{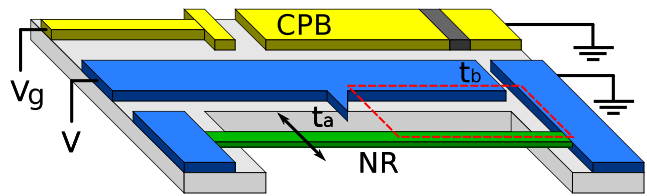}
\caption{(Color online) Possible experimental setup consisting of a qubit and an oscillator coupled to an atomic point contact (APC). Electrons tunnel at the APC ($t_a$) and a fixed tunnel junction ($t_b$), which are both biased with a voltage $V$. The area enclosed by the junctions (red dashed line) is threaded with a magnetic flux to create an Aharonov-Bohm phase. The qubit is realized as a Cooper pair box (CPB, yellow). Its state can be tuned using the gate voltage $V_g$ and it couples capacitively to both junctions. The oscillation of the nanomechanical resonator (NR, green) modulates the tunneling amplitude $t_a$. As discussed in more detail in Appendix \ref{app:phases}, this setup can be used to realize the tunneling amplitude (\ref{gamma}).}
  \label{FigScheme}
\end{figure}

In this article, we propose a system which allows the detection of entanglement between an oscillator and a qubit using an electronic measurement in an atomic point contact (APC). The electronic system is based on a tunneling contact, a readout device which is known to be quantum-limited.\cite{clerk04} We find that the measurement of the current and the symmetrized current noise in this system allows the evaluation of a criterion for entanglement\cite{rigas06} based on the density matrix of the oscillator-qubit system. This allows for the detection of entanglement in \emph{arbitrary} pure or mixed states. All elements of the proposed setup have been realized separately in different experiments. Moreover, it has been shown that the current and the noise of an APC can be measured with a high accuracy. Therefore, it should be possible to combine both elements into one functional device as schematically shown in Fig.~\ref{FigScheme} and to measure its current and noise properties.

The system we investigate consists of a nanomechanical oscillator, a qubit and a biased APC. Both the oscillator and the qubit are coupled to the APC and thus modulate its transmission coefficient. The APC consists of two electron reservoirs (``left'' and ``right'') at chemical potentials $\mu_{L,R}$ which are subject to a voltage difference $V = \mu_L - \mu_R$. The Hamiltonian of the APC reads (using units $e = \hbar = 1$)
\begin{align}
 H_{el}&= \sum_{\alpha = R,L} \sum_k (\epsilon_k + \mu_\alpha) \psi^\dag_{\alpha,k} \psi_{\alpha,k}\ , \notag \\
 H_{T} &= \gamma Y \psi^\dag_L(z=0) \psi_R(z=0) + \mathrm{h.c.}\ ,
\end{align}
where $\psi^\dag_{\alpha}(z) = (1/\sqrt{L}) \sum_k e^{-i k z} \psi^\dag_{\alpha,k}$ ($\alpha = L,R$) creates an electron at position $z$ in the left/right reservoir, respectively. As a simplification, we assume a constant density of states $\rho_0 = 1/(\pi v_F)$ where $v_F$ is the Fermi velocity. In the tunneling Hamiltonian $H_T$, the counting operator $Y$ ($Y^\dag$) decreases (increases) the transferred charge by one, $Y\ket{n_R+1} = \ket{n_R}$. The corresponding number operator is defined by $\hn_R\ket{n_R} = n_R \ket{n_R}$. The tunneling amplitude $\gamma$ will be specified shortly.

The oscillator and the qubit are described by the Hamiltonians
\begin{align}
 H_R   &= \frac{p^2}{2m} + \frac{1}{2} m \Omega^2 x^2\ , \notag \\
 H_Q   &= \epsilon \sigma_z + \Delta \sigma_x\ ,
\end{align}
where $m$ and $\Omega$ denote the effective mass and the frequency of the oscillator, respectively. In the qubit Hamiltonian, $\sigma_{x,y,z}$ denote the Pauli matrices. For $\Delta = 0$, the energy difference between the two qubit states is given by $2\epsilon$. A finite $\Delta$ enables tunneling between the states.

The state of the qubit-oscillator system modulates the tunneling amplitude $\gamma$ of the APC. If the oscillator acts as one of the electron reservoirs of the APC\cite{flowers-jacobs07} as shown in Fig.~\ref{FigScheme}, the tunneling gap depends on the oscillator displacement $x$. For small $x$ one obtains $\gamma \propto \gamma_0 + \gamma_1 x$. The same dependence can also be realized for capacitive coupling.\cite{poggio08} The qubit can be realized as a Cooper pair box in which case a depletion of the electron reservoirs of the APC depending on the state of the qubit leads to an additional term $\gamma_2 \sigma_z$ in the tunneling amplitude. Irrespective of the concrete realization, to lowest order the combined effect of the oscillator and the qubit leads to
\begin{align}\label{gamma}
 \gamma = \gamma_0 + \gamma_1 x + \gamma_2 \sigma_z\ .
\end{align}
In general, the amplitudes $\gamma_j = |\gamma_j| e^{i \delta_j}$ ($j = 0,1,2$) can be complex. Since the global phase is irrelevant, we set $\delta_0 = 0$. Finite phases $\delta_{1,2}$ can be realized experimentally by closing the electric circuit using an additional tunnel junction as shown in Fig.~\ref{FigScheme}.\cite{doiron08} Threading the loop with a magnetic flux causes Aharonov-Bohm phases which can be absorbed in the tunneling amplitudes and generally lead to finite phases $\delta_1$ and $\delta_2$. This is discussed in more detail in Appendix \ref{app:phases}. The benefits of a controllable $\delta_1$ have been investigated for a system consisting of an APC and an oscillator:\cite{doiron08} while for $\delta_1 = 0$, the current noise only depends on the oscillator position $\expcts{x^2}$, a finite $\delta_1$ leads to terms proportional to $\expcts{p^2}$ and thus contains information about the oscillator momentum. Similarly, the presence of tunable phases $\delta_{1,2}$ increases the number of measurable oscillator and qubit properties.

The current operator is defined by $I = \tfrac{d}{dt} \hn_R$. Our main objective will be the calculation of the average current $\expcts{I}$ and the symmetrized noise spectral density,
\begin{align}
 S(\omega) = \frac{1}{2} \int_{-\infty}^\infty d\tau\ e^{i \omega \tau} \cumu{\left\{ I(\tau), I(0) \right\}}\ ,
\end{align}
where $\cumu{ I(\tau) I(0)} = \expct{ I(\tau) I(0)} - \expct{I(\tau)} \expct{I(0)}$ and $\{\cdot,\cdot\}$ denotes the anticommutator. We shall find that the average current as well as the noise depend on expectation values of products of qubit and oscillator operators. From these expectation values, the bipartite expectation value matrix (EVM) can be constructed. It is a complex $6 \times 6$-matrix defined by\cite{rigas06}
\begin{align}\label{chi}
 \chi = \left(\begin{matrix}
         \expct{\ket{\uparrow}\bra{\uparrow} \otimes B} & \expct{\ket{\uparrow}\bra{\downarrow} \otimes B} \\
         \expct{\ket{\downarrow}\bra{\uparrow} \otimes B} & \expct{\ket{\downarrow}\bra{\downarrow} \otimes B} \\
        \end{matrix}\right)\ ,
\end{align}
where
\begin{align}
 B = \left(\begin{matrix}
            1 & x     & p\\
	    x & x^2   & S_{xp} \\
	    p & S_{xp} & p^2
           \end{matrix}\right)
\end{align}
and we used $S_{xp} = \tfrac{1}{2} \left( xp + px \right)$. All expectation values are taken with respect to the qubit-oscillator state described by the density matrix $\rho_\text{q,osc}$, e.g.~$\expcts{\ket{\uparrow}\bra{\uparrow} x} = \Tr [ \rho_\text{q,osc} \ket{\uparrow}\bra{\uparrow} x ]$. It has been shown\cite{rigas06} that for any separable state $\rho_\text{q,osc}$,
\begin{align}\label{chi_ineq}
 \chi \pm \frac{i}{2} \rho_q \otimes \left(\begin{matrix}
                                            0 & 0 & 0 \\
					    0 & 0 & -1 \\
					    0 & 1 & 0 \\
                                           \end{matrix} \right) \geq 0\ ,
\end{align}
i.e.~both matrices must be positive semidefinite. Here, $\rho_q = \Tr_\text{osc} \rho_\text{q,osc}$ denotes the reduced density matrix of the qubit. Once the complete EVM is known, a violation of Eq.~(\ref{chi_ineq}) proves that $\rho_\text{q,osc}$ is an entangled state. In the following, we show that the current and its noise contain enough information to construct the EVM and thus deduce entanglement of the oscillator and the qubit. Hence, this measurement provides a separability criterion for an oscillator and a qubit, comparable to a Bell inequality measurement which provides a separability criterion for two qubits.

Since the oscillator-qubit state is not necessarily stationary, both the average current and the noise spectrum will in general be time dependent. Therefore, we relate the current and noise at time $t$ to the expectation values of the qubit-oscillator system taken at the same time. We calculate the current using perturbation theory in the tunneling Hamiltonian $H_T$ and using $H_0 = H_{el} + H_R + H_Q$ as the unperturbed Hamiltonian. The method is discussed in more detail in Appendix \ref{app:current}.

Calculating the time-dependence of the qubit and oscillator operators using the Hamiltonian $H_0$, the Kubo formula straightforwardly yields the average current,
\begin{align}\label{current}
 \expct{I}
&=
 2 V T_0 + \frac{V}{2} T_2 + \frac{V}{2} T_1 \expct{\hat{x}^2} \notag \\
&+
 \sqrt{T_0 T_1} \left[ 2 V \cos(\delta_{1}) \expct{\hat{x}} + \Omega \sin(\delta_{1}) \expct{\hat{p}} \right] \notag \\
&+
 2 \sqrt{T_0 T_2} \left[ V \cos(\delta_{2}) \expct{\sigma_z} + \Delta \sin(\delta_{2}) \expct{\sigma_y} \right] \notag \\
&+
 \sqrt{T_1 T_2} \Big[ V \cos(\delta_{21}) \expct{\hat{x}\sigma_z} - \tfrac{\Omega}{2} \sin(\delta_{21}) \expct{\hat{p}\sigma_z} \notag \\
&+ \Delta\sin(\delta_{21}) \expct{\hat{x} \sigma_y} \Big] + I_{0,\text{osc}} + I_{0,\text{q}}\ ,
\end{align}
where we used $\delta_{21} = \delta_2 - \delta_1$. The dimensionless conductance of the bare tunnel junction is given by $T_0 = \pi \rho_0^2 |\gamma_0|^2$. Moreover, using the relaxation rates for the oscillator and the qubit,\cite{korotkov01_2,doiron08}
\begin{align}
 \Gamma_1 &= 2 \pi \rho_0^2 |\gamma_1|^2/m\ , \notag \\
 \Gamma_2 &= 4 \pi \rho_0^2 V |\gamma_2|^2\ ,
\end{align}
we defined dimensionless conductances for the oscillator and the qubit by $T_1 = \Gamma_1/\Omega$ and $T_2 = \Gamma_2/V$. The oscillator displacement is measured in units of its zero-point motion $x_0 = 1/\sqrt{2 m \Omega}$, i.e.~$\hat{x} = x/x_0$ and $\hat{p} = 2 x_0 p$. Finally, $I_{0,\text{osc}} = -\tfrac{1}{2} T_1 \mathrm{sgn}(V) \min( |V|, \Omega) $ is the current due to the zero-point fluctuations of the oscillator. A similar term arises for the qubit, $I_{0,\text{q}}(|V| > 2 \alpha) = T_2 \Delta \expct{\sigma_x}{} \mathrm{sgn}(V)$ while $I_{0,\text{qubit}}(|V|<2\alpha) = T_2 V \expct{\sigma_x} \Delta/(2 \alpha)$ where $\alpha = \sqrt{\epsilon^2 + \Delta^2}$.

For $\gamma_2 = 0$, the result coincides with the known result for a system containing only the oscillator coupled to an APC,\cite{doiron08} while for $\gamma_1=\delta_2=0$, the known result for an APC coupled to a qubit emerges.\cite{korotkov01_2} In the general case, a current measurement enables us to deduce the qubit expectation values needed for the construction of the reduced density matrix $\rho_q$. However, the correlation functions contained in the result for the current are insufficient to construct the EVM.

The calculation of the noise spectral density can most easily be accomplished by using the Redfield equation in connection with the Born-Markov approximation. For this purpose, we split the complete system into a fermionic ``bath'', $H_B = H_{el}$ and the actual ``system'', $H_S = H_Q + H_R$, coupled by $H_T$. The bath timescale can be estimated to be of order $1/V$ since $eV/(2 \pi \hbar)$ corresponds to the attempt frequency at which electrons arrive at the tunnel junction. In the limit $V \gg \Omega,\Delta,\epsilon$, we can assume the bath timescale to be much faster than the system timescales. To lowest nonvanishing order in the tunneling, the equation of motion for the reduced system density matrix $\rho_S(t) = \Tr_B \rho(t)$ then reads
\begin{align}\label{redfield}
 \dot{\rho}_S(t)
&= -i [ H_S, \rho_S(t) ] \notag \\
&- \int_0^\infty d\tau \Tr_B [ H_T, [ H_T(-\tau), \rho_S(t) \otimes \rho_B ] ]\ .
\end{align}
This equation can be used to calculate expectation values of system operators. In particular, the average current can be calculated using $\expcts{I} = \tfrac{d}{dt} \expct{\hn_R} = \Tr_S [ \hn_R \dot{\rho}_S ]$. The current calculated in this way reproduces Eq.~(\ref{current}), which was derived without using the Born-Markov approximation, as soon as $V > \Omega,2 \sqrt{\epsilon^2 + \Delta^2}$.

In order to use Eq.~(\ref{redfield}) to calculate the symmetrized noise, we make use of MacDonald's formula,\cite{macdonald49}
\begin{align}\label{macdonald}
 S(\omega) =  \omega \int_0^\infty dt\ \sin(\omega t) \frac{d}{dt} \cumu{\hn_R^2(t)}\ .
\end{align}
The calculation of the cumulant derivative $\tfrac{d}{dt} \cumu{\hn_R^2(t)}$ using Eq.~(\ref{redfield}) leads to expressions containing cumulants $\cumu{\hn_R X}$, where $X$ is in general a product of system operators. The time evolution of these cumulants will lead to ever higher-order cumulants, such that the resulting set of differential equations is not closed. In order to make the problem tractable, several approximations are needed.

Since the qubit and the oscillator are only weakly coupled to the APC, we can assume $\gamma_1 x_0, \gamma_2 \ll \gamma_0$. In this case, we can ignore higher-order cumulants of the form $\cumus{\hn_R x^2}, \cumus{\hn_R x \sigma_j}$, etc. Setting these higher-order cumulants to zero is often referred to as the \emph{Gaussian} approximation.

Second, the Redfield equation leads to couplings between qubit and oscillator; e.g.,~$\tfrac{d}{dt} \cumu{\hn_R x}$ contains a term depending on $\cumu{\hn_R \sigma_z}$. However, since such crossterms are subleading in terms of tunneling amplitudes, we can neglect them and thereby decouple the set of differential equations into qubit and oscillator parts.

Finally, we set $\epsilon =0$ and $\delta_2 = 0$. This is not a crucial approximation but it simplifies the differential equations considerably.

\begin{figure}[t]
  \centering
  \includegraphics[width = 0.47\textwidth]{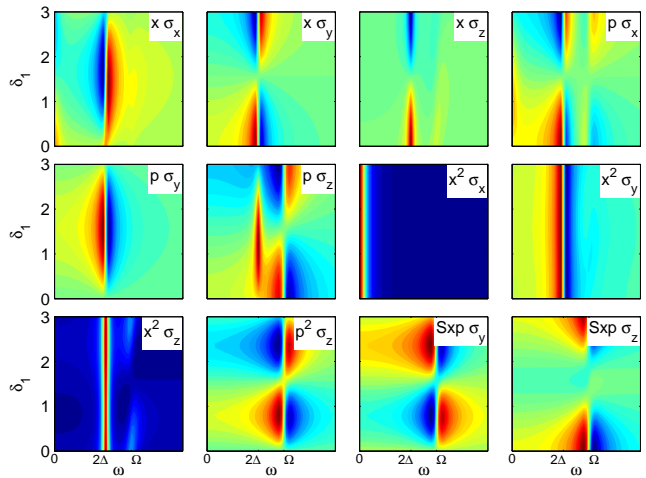}
\caption{(Color online) Schematic density plot of the prefactors $S_X(\omega)$ $(X = x\sigma_x, x\sigma_y,\ldots)$ of the frequency-dependent noise $S(\omega) = \sum_X S_X(\omega) \expct{X}$ as a function of $\delta_1$ and $\omega$.}
  \label{NoisePlot}
\end{figure}

It turns out that this set of approximations makes a solution of the Redfield equation possible and the noise can be calculated using Eq.~(\ref{macdonald}). For details, we refer to Appendix \ref{app:noise}. The resulting expression $S(\omega)$ contains all entries of the EVM with the exception of $\expct{p^2 \sigma_x}{}, \expct{p^2 \sigma_y}{}$ and $\expct{S_{xp} \sigma_x}{}$. The frequency-dependence is characterized by the following functions,
\begin{align}
 \alpha_1(\omega) &= \frac{\Gamma_1 \omega^2 \sin(\delta_1) - 2 V (\omega^2 - \Omega^2) \cos(\delta_1)}{\Gamma_1^2 \omega^2 + (\omega^2 - \Omega^2)^2}\ , \notag \\
 \alpha_2(\omega) &= \frac{2 V \Gamma_1 \Omega \cos(\delta_1) + \Omega ( \omega^2 - \Omega^2) \sin(\delta_1)}{\Gamma_1^2 \omega^2 + (\omega^2 - \Omega^2)^2}\ , \notag \\
 \beta_1(\omega) &= \frac{V ( 4 \Delta^2 - \omega^2)}{\Gamma_2^2 \omega^2 + (\omega^2 - 4 \Delta^2)^2}\ , \notag \\
 \beta_2(\omega) &= \frac{\Gamma_2 \Delta^2}{\Gamma_2^2 \omega^2 + (\omega^2 - 4 \Delta^2)^2}\ , \notag \\
 \beta_3(\omega) &= \frac{\Gamma_2}{\Gamma_2^2 + \omega^2}\ ,
\end{align}
which contain Lorentz and Fano shaped resonances at the characteristic frequencies of the system, $\omega = 0, \Omega, 2\Delta$. The complete expression for the noise reads $S(\omega) = \sum_X S_X(\omega) \expct{X}$ where $X$ denotes all combinations of qubit and oscillator operators contained in the EVM (\ref{chi}). As mentioned above, it turns out that all except three of the prefactors $S_X(\omega)$ are nonvanishing and are distinguishable combinations of the functions $\alpha_{1,2}(\omega)$ and $\beta_{1,2,3}(\omega)$. Results for $S_X(\omega)$ can be found in Appendix \ref{app:noise}. A plot of the relevant cross-correlations' prefactors is shown in Fig.~\ref{NoisePlot}. Since the shapes of these functions are rather distinct, the expectation values constituting the EVM can be recovered from the total measurable noise $S(\omega)$.

Beyond the regime $V \gg \Omega,\Delta$, the Born-Markov approximation fails and we have to resort to conventional perturbation theory in order to calculate $S(\omega)$. For the case of a simple tunneling junction, it is well known that $S(\omega)$ has kinks at $|\omega| = |V|$.\cite{schoelkopf97,blanter00} In the presence of the oscillator and the qubit, we find that $S(\omega)$ shows similar features at $|\omega| = |V \pm 2\Delta|$ and $|\omega| = |V \pm \Omega|$. The amplitudes of these kinks yield additional information about the expectation values $\expct{x \sigma_j}$ and $\expct{p \sigma_j}$ ($j = x,y,z$). Details are given in Appendix \ref{app:noise}.

Even under ideal circumstances, not all coefficients of the EVM can be determined by a current and noise measurement. However, it was shown that even incomplete knowledge of the EVM allows for the detection of entanglement for experimentally relevant states.\cite{rigas06} If the EVM contains a few unknown parameters $a_i$, we can only detect entanglement for states where Eq.~(\ref{chi_ineq}) is violated for arbitrary $a_i$. Because of the special structure of the inequalities (\ref{chi_ineq}), this numerical problem reduces to a convex optimization problem which can be efficiently solved using semidefinite programming as shown in [\onlinecite{rigas06}]. A route to obtain \emph{all} matrix elements of the EVM would be to rotate the qubit state by applying $\pi/2$-pulses using a tunable gate voltage.\cite{armour02,wallraff05,lahaye09}

The detection of entanglement based on the EVM is highly versatile since it works even for arbitrary mixed states. The problem is of course greatly simplified if properties of the state to be measured are known, e.g.~due to a tailored preparation of the system. As an example, consider the qubit-oscillator state $\ket{\psi} = \sqrt{p_0} \ket{\uparrow,n} + i \sqrt{1 - p_0} \ket{\downarrow,n+1}$, where $\ket{n}$ denotes a Fock state of the oscillator and $0 \leq p_0 \leq 1$. If the system is assumed to be in a state of this form, detection of entanglement reduces to a measurement of $p_0$. For this state, it turns out that $S(\Omega,\delta_1) - S(\Omega,-\delta_1) \sim \sqrt{ p_0 (1-p_0)} \sin(\delta_1)$. A measurement of the noise at the oscillator resonance frequency for finite $\delta_1$ is then sufficient to detect entanglement. Similar relations can be derived for other states. The less is known about the state, the more information has to be gained from $S(\omega,\delta_1)$. We found that even for \emph{completely arbitrary} states, entanglement can be detected by this electronic measurement. Therefore, it is our expectation that this setup is ideally suited to detect entanglement between a continuous and a discrete quantum system.

\acknowledgments We thank N.~Flowers-Jacobs, D.~Loss, and particularly O.~G{\"u}hne for useful discussions. TLS and CB acknowledge support from the Swiss NSF and the NCCR Nanoscience. KB acknowledges support from the Research Council of Norway, Grant No.~191576/V30 (FRINAT) and BT from the German DFG.

\appendix

\section{Tunneling phases}\label{app:phases}

\begin{figure}[t]
  \centering
  \begin{tabular}{|lc|lc|}
   \hline
   (a) & & (b) &\\
   & \includegraphics{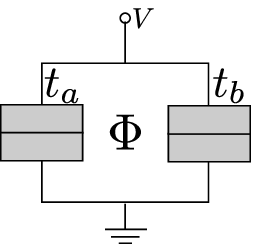} & 
   & \includegraphics{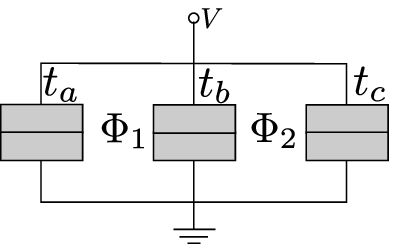} \\
   \hline
  \end{tabular}
  \caption{(a) A finite phase $\delta_1$ can be achieved by using two junctions with amplitudes $t_a(x,\sigma_z)$ and $t_b(\sigma_z)$. The phase can be controlled by tuning the magnetic flux $\Phi$. (b) A third junction $t_c$ and a two magnetic fluxes $\Phi_{1,2}$ yield control over both phases $\delta_{1,2}$.}
  \label{fig:phases}
\end{figure}

Phases of the tunneling amplitudes can be created by adding a second tunneling junction to the system and threading the resulting loop with a magnetic flux $\Phi$. If the setup of Fig.~\ref{FigScheme} is chosen, one of the tunneling amplitudes will be $x$-dependent while the other one will not. A schematic drawing is shown in Fig.~\ref{fig:phases}a, and the tunneling amplitudes of the two junctions are given by
\begin{align}
 t_a &= t_{a0} + t_{a1} \hat{x} + t_{a2} \sigma_z, \notag \\
 t_b &= t_{b0} + t_{b2} \sigma_z ,
\end{align}
where $\hat{x} = x/x_0$. Due to the Aharonov-Bohm effect, the magnetic flux leads to a phase shift in the electronic wave function which can be absorbed in the tunneling amplitudes. The total transmission amplitude is given by the sum of the two amplitudes,
\begin{align}
 t = t_a + t_b e^{-i \Phi/\Phi_0},
\end{align}
where $\Phi_0 = h/e$ is the magnetic flux quantum. We assume that the influence of the qubit state on the amplitudes is the same for both junctions. Then, the amplitudes are related by a real constant $c$, $t_{a,j} = c t_{b,j}$ for $j = 0,2$. The total tunneling amplitude now becomes (up to an irrelevant global phase)
\begin{align}
 t 
&= 
 \sqrt{1 + c^2} ( t_{a0} + t_{a2} \sigma_z ) + t_{a1} e^{-i \delta_1} \hat{x},
\end{align}
where $\delta_1 = \arg(1 + c e^{-i \Phi/\Phi_0})$. Hence, this setup provides a way to obtain a tunneling Hamiltonian with tunable $\delta_1$. This has been important for the calculation of the noise. The setup can easily be extended to achieve the second tunable phase $\delta_2$ which we used in the calculation of the current. Here, we need a third junction with an amplitude $t_c = t_{c0}$, which is decoupled from both oscillator and qubit, and two magnetic fluxes $\Phi_{1,2}$. A schematic is shown in Fig.~\ref{fig:phases}b.

Note that in order to observe Aharonov-Bohm phases, the phase coherence length (which can be on the order of microns) must exceed the size of the setup. Moreover, the change of the area due to the fluctuating position of the resonator is assumed to be negligible. Our analysis only covers the single-channel case. In the case of $N$ channels in the loop, the magnitude of the Aharonov-Bohm effect is reduced by a factor $1/N$. Finally, we assume that roundtrips of electrons are ruled out by the device geometry such that electrons leave the setup after passing through either of the junctions. In this case, one can generally have $|t(\Phi)|^2 \neq|t(-\Phi)|^2$.

\section{Current calculation}\label{app:current}

The current operator is $I = \frac{d}{dt} \hat{n}_R = i (A \psi_L^\dag \psi_R - A^\dag \psi_R^\dag \psi_L)$, where $A = \gamma Y$ and $\psi_\alpha \equiv \psi_\alpha(z=0)$. The tunneling amplitude $\gamma$ is defined in Eq.~(3). The average current is in general time-dependent, and can be calculated by treating $H_T$ as a perturbation to the non-interacting Hamiltonian $H_0 = H_{el} + H_R + H_Q$. In that case, the Kubo formula gives
\begin{equation}
  \label{eq:KuboCurrent}
  \langle I(t) \rangle  = -i \int_0^t dt' \, \langle \Psi(0) | \left[I_{\mathrm{int}}(t) \, , \, H_{T,\mathrm{int}}(t') \right] | \Psi(0) \rangle ,
\end{equation}
where the subscript ``int'' refers to the interaction picture and $|\Psi(0)\rangle$ is the state of the total system at time $t = 0$. We shall assume that $|\Psi(0)\rangle$ is a product state of the fermionic ground state and the (possibly entangled) qubit-oscillator state.

The current can also be expressed in terms of the state at time $t$, which {\it to zeroth order} in the tunneling amplitudes $\gamma_j$ is $|\Psi(t)\rangle = e^{i H_0 t} |\Psi(0)\rangle$. Thus, to second order in the tunneling amplitudes, the current becomes
\begin{align}
\langle I(t) \rangle &= 2 \, \mathrm{Re} \, \int_0^t d\tau \notag \\
 & \langle \Psi(t) | [A \psi^\dag_L \psi_R \, , \, (A^\dag \psi_R^\dag \psi_L)_{\mathrm{int}}(-\tau) ] | \Psi(t) \rangle  .
\end{align}
If the time $t$ is large compared to the timescales of the non-interacting system [$V^{-1},\Omega^{-1}, (\epsilon^2+\Delta^2)^{-1/2}$], we can extend the $\tau$-integral to infinity. On the other hand, the above approximations are only valid for $t$ smaller than any damping time scale, both due to the coupling to the leads ($\sim \, \Gamma_1^{-1},\Gamma_2^{-1}$) and to other sources of dissipation not included in our model. \\

Next, we define the electronic correlation functions
\begin{align}\label{Z}
 Z_B^>(-\tau,V) &= \expcts{\psi_L^\dag(0) \psi_L(-\tau)}{} \expcts{\psi_R(0) \psi_R^\dag(-\tau)}{} \notag \\
 Z_B^<(-\tau,V) &= \expcts{\psi_L(-\tau) \psi_L^\dag(0)}{} \expcts{\psi_R^\dag(-\tau) \psi_R(0)}{} .
\end{align}
The averages are taken with respect to the non-interacting ground state such that these correlation functions can easily be calculated. The chemical potentials of the left and right reservoirs differ by the applied voltage $V$. For simplicity, we focus on the simplest case of zero temperature and a constant density of states $\rho_0$. For zero voltage, one finds after Fourier transformation
\begin{align}\label{ZFourierT0}
 Z_B^>(\omega) \equiv Z_B^>(\omega, V=0) &= - 2 \pi \rho_0^2 \theta(- \omega) \omega \notag \\
 Z_B^<(\omega) \equiv Z_B^<(\omega, V=0)  &= 2 \pi \rho_0^2 \theta(\omega) \omega.
\end{align}
A finite voltage leads to additional phase factors $Z_B^>(\tau,V) = e^{-i V \tau} Z_B^>(\tau)$. By exploiting $Y Y^\dagger = 1$, one then obtains the following expression for the average current
\begin{align}
  \label{eq:CurrentwithZ}
  \langle I(t) \rangle &=  2 \, \mathrm{Re} \, \int_0^\infty d\tau \, e^{i V \tau} \Big(Z^>_B(-\tau) \, \langle \gamma \, \gamma^\dagger(-\tau)  \rangle \notag \\
&-  Z^<_B(-\tau) \, \langle  \gamma^\dagger(-\tau) \, \gamma \rangle \Big)  ,
\end{align}
where the expectation value is in the state at time $t$. The time dependence of the operator $\gamma$ is in the interaction picture and is given by the free evolution of the oscillator and the qubit,
\begin{align}
  \label{eq:FreeEvolution}
  x(t) &= x \cos(\Omega t) + \frac{p}{m \Omega} \sin(\Omega t) , \notag \\
 \sigma_z(t)
&=
 \frac{\epsilon^2 + \Delta^2 \cos(2 \sqrt{\epsilon^2 + \Delta^2} t)}{\epsilon^2 + \Delta^2} \sigma_z \notag \\
&+ \frac{\Delta \epsilon [ 1 - \cos(2 \sqrt{\epsilon^2 + \Delta^2} t) ]}{\epsilon^2 + \Delta^2} \sigma_x \notag \\
&+ \frac{\Delta \sin(2 \sqrt{\epsilon^2 + \Delta^2} t)}{\sqrt{\epsilon^2 + \Delta^2}} \sigma_y .
\end{align}
The final expression (8) for the current can be obtained by using this, as well as $[x,p] = i$, $\sigma_i \sigma_j = \delta_{ij} + i \varepsilon_{ijk} \sigma_k$, and the assumption that the cut-off frequency of the density of states in the leads is much larger than $V,\Omega,\sqrt{\epsilon^2+\Delta^2}$.

In conclusion, given the state $|\Psi(0)\rangle$ at $t=0$ assumed above, the expression for the average current is valid for times $ V^{-1},\Omega^{-1},(\epsilon^2+\Delta^2)^{-1/2} \ll t \ll \Gamma_1^{-1},\Gamma_2^{-1}$.

\section{Noise calculation}\label{app:noise}

We now provide some details on the calculation of the symmetrised noise spectral density
\begin{equation}
  \label{eq:NoiseSpectralDensity}
  S(\omega,t) = \frac{1}{2} \int_{-\infty}^\infty d \tau \ e^{i \omega \tau} \langle\!\langle \left\{ I(\tau) \, , \, I(0) \right\} \rangle\!\rangle,
\end{equation}
where $\langle\!\langle I(\tau) I(0)  \rangle\!\rangle = \langle I(\tau) I(0) \rangle - \langle I(\tau) \rangle \langle I(0) \rangle$ and $\{ \cdot, \cdot \}$ denotes the anti-commutator. The time dependence is in the Heisenberg picture and the expectation values are taken in the state $|\Psi(t)\rangle$ at time $t$.

\subsection{Perturbation theory}

To lowest order in the tunneling amplitudes, the noise spectral density can be approximated by
\begin{equation}
  \label{eq:NoiseSpectralDensityLowestOrder}
   S(\omega,t) = \frac{1}{2} \int_{-\infty}^\infty d \tau \ e^{i \omega \tau} \, \langle \Psi(t)|\left\{ I_\mathrm{int}(\tau) \, , \, I(0) \right\} |\Psi(t)\rangle,
\end{equation}
where as before $I_\mathrm{int}(\tau) = e^{iH_0\tau}\,  I \, e^{-iH_0 \tau}$. Applying the same approximations as in the calculation of the current, we find $S(\omega,t) = (\tilde{S}(\omega,t) + \tilde{S}^\ast(-\omega,t))/2$, where
\begin{align}
  \label{eq:Stilde}
  \tilde{S}(\omega,t) &= \int_{-\infty}^\infty d \tau \ e^{i (\omega + V) \tau} \Big(Z_B^>(-\tau) \langle \gamma(\tau) \gamma^\dagger \rangle \notag \\
&+ Z_B^<(-\tau) \langle \gamma^\dagger(-\tau) \gamma \rangle \Big) .
\end{align}
$S(\omega,t)$ is an even function of $\omega$. For $|\omega|,\Omega,\Delta \ll |V|$, it is frequency-independent and proportional to $\sgn(V) \langle I(t) \rangle$, which is due to electron shot noise. Assuming $\epsilon = 0$ and defining the function $f(\omega,V) = |\omega| + \Theta(|V|-|\omega|)(|V|-|\omega|)$, the noise spectral density becomes
\begin{widetext}
\begin{align}
  \label{eq:2ndOrderNoise}
  S(\omega,t) & = \pi \rho_0^2 \notag \\
& \times \Big[2  |\gamma_0| \Big( |\gamma_0| + |\gamma_1| \cos \delta_1 \langle x \rangle  + |\gamma_2| \cos \delta_2 \langle \sigma_z \rangle \Big) f(\omega,V) - 2 \Omega |\gamma_1|^2 x_0^2 + 4 \Delta |\gamma_2|^2 \langle \sigma_x \rangle \notag \\
& + |\gamma_1| \Big(|\gamma_0| \cos \delta_1 \langle x \rangle + |\gamma_1| \langle x^2 \rangle + |\gamma_2| \cos \delta_{21} \langle x \sigma_z \rangle \Big)\Big(f(\omega,V+\Omega) + f(\omega,V-\Omega)\Big) \notag \\
& + 2 x_0^2 |\gamma_1| \Big(|\gamma_0| \sin \delta_1 \langle p \rangle - |\gamma_2| \sin \delta_{21} \langle p \sigma_z \rangle \Big) \Big(f(\omega,V+\Omega) -  f(\omega,V-\Omega)\Big) \notag \\
& + |\gamma_2|\Big(|\gamma_0| \cos \delta_2 \langle \sigma_z \rangle + |\gamma_2| + |\gamma_1| \cos \delta_{21} \langle x \sigma_z \rangle \Big) \Big(f(\omega,V+2\Delta) + f(\omega,V-2\Delta)\Big) \notag \\
& + |\gamma_2| \Big(|\gamma_0| \sin \delta_2 \langle \sigma_y \rangle  + |\gamma_1| \sin \delta_{21} \langle x \sigma_y \rangle \Big) \Big(f(\omega,V+2\Delta) -  f(\omega,V-2\Delta)\Big) \Big].
\end{align}
\end{widetext}
We observe that $S(\omega,t)$ has kinks at $|\omega| = |V|$, $|\omega| = |V \pm \Omega|$ and $|\omega| = |V \pm 2\Delta|$, which can also provide information on the bipartite expectation value matrix of the oscillator-qubit system.\\

As determining higher-order contributions to the noise spectral density is complicated using this approach, we now turn to an alternative method.

\subsection{Born-Markov approximation}

In the case $V \gg \Omega, \Delta, \epsilon$, the timescale of electron dynamics will be faster than the oscillator and qubit timescales. The electron reservoirs can then be regarded as a fermionic bath $H_B = H_{el}$, coupled to the remaining system $H_S = H_R + H_Q$ via the tunneling operator $H_T$. For small tunneling, the dynamics of the reduced density matrix $\rho_S(t) = \Tr_{B} \rho(t)$, will be determined by the Redfield equation,
\begin{align}\label{Aredfield}
 \dot{\rho}_S(t)
&= -i [ H_S, \rho_S(t) ] \notag \\
&- \int_0^\infty d\tau \Tr_B [ H_T, [ H_T(-\tau), \rho_S(t) \otimes \rho_B ] ] .
\end{align}
The time-evolution of $H_T$ is taken with respect to the uncoupled Hamiltonian $H_S + H_B$. Expectation values of system operators can in principle be calculated by using e.g.~$\frac{d}{dt} \expct{x(t)}{} = \Tr_S[ \dot{\rho}_S(t) x]$ and solving the resulting differential equation. Splitting the tunnel Hamiltonian into a system and a bath part, $H_T = A \psi^\dag_L \psi_R + \hc$, we can rewrite this as
\begin{align}\label{redfield2}
  \dot{\rho}_S(t) &= -i \left[ H_S, \rho_S(t) \right] \notag \\
&-  \int_0^\infty d\tau \Big\{ Z_B^>(-\tau,V) [ A, A^\dag(-\tau) \rho_S(t)] \notag \\
&- Z_B^<(-\tau,V) [ A, \rho_S(t) A^\dag(-\tau) ] + \hc \Big\} ,
\end{align}
where we used the bath correlation functions defined in Eq.~(\ref{Z}) and the free evolution of the qubit-oscillator operators is given in Eq.~(\ref{eq:FreeEvolution}). Therefore, $A(t)$ contains terms oscillating at the frequencies $\Omega$ and $2 \sqrt{\epsilon^2 + \Delta^2}$. The $\tau$-integration in Eq.~(\ref{redfield2}) can then be performed and its result can be absorbed into generalized transmission probabilities $\Gamma^{\rightleftarrows}_{\hk\hj}$ where $\hk$ and $\hj$ run over the system operators, $\hk,\hj \in \{ x, p, \sigma_x, \sigma_y, \sigma_z \}$. The average current is defined as $\expct{I(t)}{} = \Tr_S [ \dot{\rho}_S(t) n_R ]$. Using the Redfield equation, it can be written as
\begin{align}
 \expct{I}{}
=& -
 \sum_{\hk \hj} \Big\{
   \Gamma^\rightarrow_{\hk\hj}  \expct{[n_R, Y \hk] Y^\dag \hj }{}\notag \\
-&
   \Gamma^\leftarrow_{\hk\hj} \expct{Y^\dag \hj [n_R, Y \hk]}{}
+
   \text{c.c.} \Big\} \notag \\
=&
 \sum_{\hk \hj} \Big\{
   \Gamma^\rightarrow_{\hk\hj}  \expct{\hk \hj }{}
-
   \Gamma^\leftarrow_{\hk\hj} \expct{\hj \hk}{}
+
   \text{c.c.} \Big\},
\end{align}
where we evaluated the expression using $[Y, n_R] = Y$ and $Y Y^\dag = Y^\dag Y = \mathbbm{1}$ while all other commutators of $n_R$ and $Y$ with the system operators $K, J$ vanish. The calculation of the transmission coefficients is a straightforward task and the average current for $V > \Omega, \sqrt{\epsilon^2 + \Delta^2}$ coincides exactly with the result from the perturbation theory.

In the following, we consider the case $\epsilon = \delta_2 = 0$. The symmetrized frequency-dependent current noise is defined by Eq.~(\ref{eq:NoiseSpectralDensity}). In order to transform this into an expression which can be calculated using the Redfield equation, we use MacDonald's formula, see Eq.~(11). The time-derivative of the cumulant of the number of transfered electrons $\cumus{n_R^2} = \expcts{n_R^2} - \expct{n_R}^2$ can be calculated using Eq.~(\ref{redfield2}). The noise turns out to have a zero-frequency component proportional to the average current $S(\omega = 0) = 2e\expct{I}{}$. After subtracting this term, one finds
\begin{widetext}
\begin{align}
 \frac{d}{dt} \cumu{n_R^2} - \expct{I}
&=
 \frac{d}{dt} \expct{n_R^2}{} - 2 \expct{n_R}{} \expct{I}{} - \expct{I}{} \notag \\
&=
 4 V \sqrt{\frac{\Gamma_0 \Gamma_1}{\Omega}} \cos(\delta_1) \cumu{n_R \hat{x}} +
 2 \sqrt{\Gamma_0 \Gamma_1 \Omega} \sin(\delta_1) \cumu{n_R \hat{p}} +
 \frac{V \Gamma_1}{\Omega} \cumu{n_R \hat{x}^2} \notag \\
&+
 \frac{2 \Delta \Gamma_2}{V} \cumu{n_R \sigma_x} +
 4 \sqrt{V \Gamma_0 \Gamma_2} \cumu{n_R \sigma_z} \notag \\
&+
 2 \sqrt{\frac{V \Gamma_1 \Gamma_2}{\Omega}} \cos(\delta_1) \cumu{n_R \hat{x} \sigma_z} -
 2 \Delta \sqrt{\frac{\Gamma_1 \Gamma_2}{V \Omega}} \sin(\delta_1) \cumu{n_R \hat{x} \sigma_y} +
 \sqrt{\frac{\Gamma_1 \Gamma_2 \Omega}{V}} \sin(\delta_1) \cumu{n_R \hat{p} \sigma_z},
\end{align}
\end{widetext}
where the oscillator coordinate was measured in units of its zero-point motion, $\hat{x} = x/x_0$, $\hat{p} = 2 x_0 p$ where $x_0 = 1/\sqrt{2 m \Omega}$. Next, we assume that the bare tunneling amplitude is large compared to the coupling to the qubit and the oscillator, ie.~$\gamma_0 \gg \gamma_1 x_0, \gamma_2$. Since all of the above cumulants vanish in the absense of coupling to the qubit and the oscillator, they have to be on the order of these couplings $\gamma_1, \gamma_2$. Therefore, cumulants containing quadratic operators in addition to $n_R$ are subleading and we can use
\begin{align}
 \frac{d}{dt} \cumu{n_R^2} - \expct{I}
&=
 4 V \sqrt{\frac{\Gamma_0 \Gamma_1}{\Omega}} \cos(\delta_1) \cumu{n_R \hat{x}}  \\
&+
 2 \sqrt{\Gamma_0 \Gamma_1 \Omega} \sin(\delta_1) \cumu{n_R \hat{p}} \notag \\
&+
 \frac{2 \Delta \Gamma_2}{V} \cumu{n_R \sigma_x} +
 4 \sqrt{V \Gamma_0 \Gamma_2} \cumu{n_R \sigma_z}. \notag
\end{align}
Since for a Gaussian distribution, all cumulants beyond quadratic order vanish, we call this the Gaussian approximation. The next step is to calculate the time-dependence of the cumulants which can again be accomplished using the Redfield equation (\ref{redfield2}). In order to simplify the expressions, we can assume $\expcts{x}{} = \expcts{p}{} = \expcts{\sigma_z}{} = 0$ since finite values for these expectation values would merely lead to a renormalized tunneling amplitude. As an example, consider the differential equation for $\cumus{n_R p}$,
\begin{align}
 \frac{d}{dt} \cumu{n_R p}
&=
 (\cdots) -
 2 \Delta \sqrt{\frac{\Gamma_1 \Gamma_2}{V \Omega}} \cos(\delta_1) \cumu{n_R \sigma_y} \notag \\
&-
 \sqrt{\frac{V \Gamma_1\Gamma_2}{\Omega}} \sin(\delta_1) \cumu{n_R \sigma_z} \notag \\
&-
 \Omega \cumu{n_R x} -
 \Gamma_1 \cumu{n_R p},
\end{align}
where $(\cdots)$ denotes a number of $n_R$-independent terms and the initial condition reads $\cumu{n_R p}_{t = 0}=0$. For the reason mentioned above, the terms containing cumulants of qubit operators are subleading and can be dropped. The same is true for the remaining differential equation, such that equations describing qubit cumulants and oscillator cumulants decouple. This makes an analytical solution of the problem possible. The result is an expression for the noise of the form $S(\omega) = \sum_X S_X(\omega) \expct{X}{}$, where $X$ denotes qubit or oscillator operators or products of these. The prefactors $S_X(\omega)$ for the cross-terms turn out to be given by
\begin{widetext}
\begin{align}\label{noise_prefactors}
 S_{x \sigma_x} &= 2 \sqrt{2 M \Gamma_0 \Gamma_1} \Gamma_2 (\Delta/V) \left[ \Omega \alpha_2(\omega) + 2 V \beta_3(\omega) \cos(\delta_1) + 8 V \beta_2(\omega) \cos(\delta_1) - 4 V \beta_1(\omega) \sin(\delta_1) \right]  \notag \\
 S_{x \sigma_y} &= 2 \sqrt{2 M \Gamma_1 \Gamma_2 / V} \Delta \left[ 8 V \Gamma_0 \beta_1(\omega) \cos(\delta_1) + \Gamma_2 \beta_3(\omega) \sin(\delta_1) \right]   \notag \\
 S_{x \sigma_z} &= 2 \sqrt{2 M V \Gamma_1 \Gamma_2}  \left[ - 2 \Gamma_2 (\Delta/V)^2 \beta_3(\omega) \cos(\delta_1) + 16 V \Gamma_0 \beta_2(\omega) \cos(\delta_1) + 2 \Gamma_0 \Omega \alpha_2(\omega) \right]   \notag \\
 S_{p \sigma_x} &= 2 \sqrt{2 \Gamma_0 \Gamma_1 / M} \Gamma_2 (\Delta/V) \left[ \alpha_1(\omega) + 2 \beta_1(\omega)  \cos(\delta_1)  + \beta_3(\omega) \sin(\delta_1) \right]  \notag \\
 S_{p \sigma_y} &= - \sqrt{2 \Gamma_1 \Gamma_2/(VM)} (\Delta/V) \left[ \Gamma_2 \beta_3(\omega) \cos(\delta_1) - 8 V \Gamma_0 \beta_1(\omega) \sin(\delta_1) \right]   \notag \\
 S_{p \sigma_z} &= 4 \sqrt{ 2 V \Gamma_1 \Gamma_2 / M} \Gamma_0 \left[ \alpha_1(\omega) + 4 \beta_2(\omega) \sin(\delta_1) \right]  \notag \\
 S_{x^2 \sigma_x} &= 2 M \Gamma_1 \Gamma_2 \Delta \beta_3(\omega)  \notag \\
 S_{x^2 \sigma_y} &= 4 M \sqrt{\Gamma_0 \Gamma_2/V} \Gamma_1 \Delta \left[ -\Omega \alpha_2(\omega) \sin(\delta_1) + 2 V \beta_1(\omega) \right]  \notag \\
 S_{x^2 \sigma_z} &= 4 M \Gamma_1 \sqrt{V \Gamma_0 \Gamma_2} \left[ 4 V \beta_2(\omega) + \Omega \alpha_2(\omega) \cos(\delta_1) \right]  \notag \\
 S_{p^2 \sigma_x} &= 0  \notag \\
 S_{p^2 \sigma_y} &= 0  \notag \\
 S_{p^2 \sigma_z} &= 2 \sqrt{\Gamma_0 \Gamma_2/V} (\Gamma_1/M) \alpha_1(\omega) \sin(\delta_1)  \notag \\
 S_{S_{xp} \sigma_x} &= 0  \notag \\
 S_{S_{xp} \sigma_y} &= - 2 \sqrt{\Gamma_0 \Gamma_2/V} \Gamma_1 \Delta \alpha_1(\omega) \sin(\delta_1) \notag \\
 S_{S_{xp} \sigma_z} &= \sqrt{\Gamma_0 \Gamma_2/V} \Gamma_1 \left[ 2 V\alpha_1(\omega) \cos(\delta_1) + \Omega \alpha_2(\omega) \sin(\delta_1) \right],
\end{align}
\end{widetext}
where the functions $\alpha_{1,2}(\omega)$ and $\beta_{1,2,3}(\omega)$ were defined in Eq.~(12).


\begin{thebibliography}{27}
\expandafter\ifx\csname natexlab\endcsname\relax\def\natexlab#1{#1}\fi
\expandafter\ifx\csname bibnamefont\endcsname\relax
  \def\bibnamefont#1{#1}\fi
\expandafter\ifx\csname bibfnamefont\endcsname\relax
  \def\bibfnamefont#1{#1}\fi
\expandafter\ifx\csname citenamefont\endcsname\relax
  \def\citenamefont#1{#1}\fi
\expandafter\ifx\csname url\endcsname\relax
  \def\url#1{\texttt{#1}}\fi
\expandafter\ifx\csname urlprefix\endcsname\relax\def\urlprefix{URL }\fi
\providecommand{\bibinfo}[2]{#2}
\providecommand{\eprint}[2][]{\url{#2}}

\bibitem[{\citenamefont{Schwab and Roukes}(2005)}]{schwab05}
\bibinfo{author}{\bibfnamefont{K.~C.} \bibnamefont{Schwab}} \bibnamefont{and}
  \bibinfo{author}{\bibfnamefont{M.~L.} \bibnamefont{Roukes}},
  \bibinfo{journal}{Phys. Today} \textbf{\bibinfo{volume}{58}},
  \bibinfo{pages}{36} (\bibinfo{year}{2005}).

\bibitem[{\citenamefont{LaHaye et~al.}(2004)\citenamefont{LaHaye, Buu,
  Camarota, and Schwab}}]{lahaye04}
\bibinfo{author}{\bibfnamefont{M.~D.} \bibnamefont{LaHaye}},
  \bibinfo{author}{\bibfnamefont{O.}~\bibnamefont{Buu}},
  \bibinfo{author}{\bibfnamefont{B.}~\bibnamefont{Camarota}}, \bibnamefont{and}
  \bibinfo{author}{\bibfnamefont{K.~C.} \bibnamefont{Schwab}},
  \bibinfo{journal}{Science} \textbf{\bibinfo{volume}{304}},
  \bibinfo{pages}{74} (\bibinfo{year}{2004}).

\bibitem[{\citenamefont{Mamin and Rugar}(2001)}]{mamin01}
\bibinfo{author}{\bibfnamefont{H.~J.} \bibnamefont{Mamin}} \bibnamefont{and}
  \bibinfo{author}{\bibfnamefont{D.}~\bibnamefont{Rugar}},
  \bibinfo{journal}{Appl. Phys. Lett.} \textbf{\bibinfo{volume}{79}},
  \bibinfo{pages}{3358} (\bibinfo{year}{2001}).

\bibitem[{\citenamefont{Yang et~al.}(2006)\citenamefont{Yang, Callegari, Feng,
  Ekinci, and Roukes}}]{yang06}
\bibinfo{author}{\bibfnamefont{Y.~T.} \bibnamefont{Yang}},
  \bibinfo{author}{\bibfnamefont{C.}~\bibnamefont{Callegari}},
  \bibinfo{author}{\bibfnamefont{X.~L.} \bibnamefont{Feng}},
  \bibinfo{author}{\bibfnamefont{K.~L.} \bibnamefont{Ekinci}},
  \bibnamefont{and} \bibinfo{author}{\bibfnamefont{M.~L.}
  \bibnamefont{Roukes}}, \bibinfo{journal}{Nano Lett.}
  \textbf{\bibinfo{volume}{6}}, \bibinfo{pages}{583} (\bibinfo{year}{2006}).

\bibitem[{\citenamefont{LaHaye et~al.}(2009)\citenamefont{LaHaye, Suh,
  Echternach, Schwab, and Roukes}}]{lahaye09}
\bibinfo{author}{\bibfnamefont{M.~D.} \bibnamefont{LaHaye}},
  \bibinfo{author}{\bibfnamefont{J.}~\bibnamefont{Suh}},
  \bibinfo{author}{\bibfnamefont{P.~M.} \bibnamefont{Echternach}},
  \bibinfo{author}{\bibfnamefont{K.~C.} \bibnamefont{Schwab}},
  \bibnamefont{and} \bibinfo{author}{\bibfnamefont{M.~L.}
  \bibnamefont{Roukes}}, \bibinfo{journal}{Nature}
  \textbf{\bibinfo{volume}{459}}, \bibinfo{pages}{960} (\bibinfo{year}{2009}).

\bibitem[{\citenamefont{Courty et~al.}(2001)\citenamefont{Courty, Heidmann, and
  Pinard}}]{courty01}
\bibinfo{author}{\bibfnamefont{J.-M.} \bibnamefont{Courty}},
  \bibinfo{author}{\bibfnamefont{A.}~\bibnamefont{Heidmann}}, \bibnamefont{and}
  \bibinfo{author}{\bibfnamefont{M.}~\bibnamefont{Pinard}},
  \bibinfo{journal}{Eur. Phys. J. D} \textbf{\bibinfo{volume}{17}},
  \bibinfo{pages}{399} (\bibinfo{year}{2001}).

\bibitem[{\citenamefont{Wilson-Rae et~al.}(2007)\citenamefont{Wilson-Rae,
  Nooshi, Zwerger, and Kippenberg}}]{wilson-rae07}
\bibinfo{author}{\bibfnamefont{I.}~\bibnamefont{Wilson-Rae}},
  \bibinfo{author}{\bibfnamefont{N.}~\bibnamefont{Nooshi}},
  \bibinfo{author}{\bibfnamefont{W.}~\bibnamefont{Zwerger}}, \bibnamefont{and}
  \bibinfo{author}{\bibfnamefont{T.~J.} \bibnamefont{Kippenberg}},
  \bibinfo{journal}{Phys. Rev. Lett.} \textbf{\bibinfo{volume}{99}},
  \bibinfo{pages}{093901} (\bibinfo{year}{2007}).

\bibitem[{\citenamefont{Marquardt et~al.}(2007)\citenamefont{Marquardt, Chen,
  Clerk, and Girvin}}]{marquardt07}
\bibinfo{author}{\bibfnamefont{F.}~\bibnamefont{Marquardt}},
  \bibinfo{author}{\bibfnamefont{J.~P.} \bibnamefont{Chen}},
  \bibinfo{author}{\bibfnamefont{A.~A.} \bibnamefont{Clerk}}, \bibnamefont{and}
  \bibinfo{author}{\bibfnamefont{S.~M.} \bibnamefont{Girvin}},
  \bibinfo{journal}{Phys. Rev. Lett.} \textbf{\bibinfo{volume}{99}},
  \bibinfo{pages}{093902} (\bibinfo{year}{2007}).

\bibitem[{\citenamefont{Naik et~al.}(2006)\citenamefont{Naik, Buu, LaHaye,
  Armour, Clerk, Blencowe, and Schwab}}]{naik06}
\bibinfo{author}{\bibfnamefont{A.}~\bibnamefont{Naik}},
  \bibinfo{author}{\bibfnamefont{O.}~\bibnamefont{Buu}},
  \bibinfo{author}{\bibfnamefont{M.~D.} \bibnamefont{LaHaye}},
  \bibinfo{author}{\bibfnamefont{A.~D.} \bibnamefont{Armour}},
  \bibinfo{author}{\bibfnamefont{A.~A.} \bibnamefont{Clerk}},
  \bibinfo{author}{\bibfnamefont{M.~P.} \bibnamefont{Blencowe}},
  \bibnamefont{and} \bibinfo{author}{\bibfnamefont{K.~C.}
  \bibnamefont{Schwab}}, \bibinfo{journal}{Nature}
  \textbf{\bibinfo{volume}{443}}, \bibinfo{pages}{193} (\bibinfo{year}{2006}).

\bibitem[{\citenamefont{Rocheleau et~al.}(2010)\citenamefont{Rocheleau, Ndukum,
  Macklin, Hertzberg, Clerk, and Schwab}}]{rocheleau09}
\bibinfo{author}{\bibfnamefont{T.}~\bibnamefont{Rocheleau}},
  \bibinfo{author}{\bibfnamefont{T.}~\bibnamefont{Ndukum}},
  \bibinfo{author}{\bibfnamefont{C.}~\bibnamefont{Macklin}},
  \bibinfo{author}{\bibfnamefont{J.~B.} \bibnamefont{Hertzberg}},
  \bibinfo{author}{\bibfnamefont{A.~A.} \bibnamefont{Clerk}}, \bibnamefont{and}
  \bibinfo{author}{\bibfnamefont{K.~C.} \bibnamefont{Schwab}},
  \bibinfo{journal}{Nature} \textbf{\bibinfo{volume}{463}}, \bibinfo{pages}{72}
  (\bibinfo{year}{2010}).

\bibitem[{\citenamefont{H\"uttel et~al.}(2009)\citenamefont{H\"uttel, Steele,
  Witkamp, Poot, Kouwenhoven, and van~der Zant}}]{huettel09}
\bibinfo{author}{\bibfnamefont{A.~K.} \bibnamefont{H\"uttel}},
  \bibinfo{author}{\bibfnamefont{G.~A.} \bibnamefont{Steele}},
  \bibinfo{author}{\bibfnamefont{B.}~\bibnamefont{Witkamp}},
  \bibinfo{author}{\bibfnamefont{M.}~\bibnamefont{Poot}},
  \bibinfo{author}{\bibfnamefont{L.~P.} \bibnamefont{Kouwenhoven}},
  \bibnamefont{and} \bibinfo{author}{\bibfnamefont{H.~S.~J.}
  \bibnamefont{van~der Zant}}, \bibinfo{journal}{Nano Lett.}
  \textbf{\bibinfo{volume}{9}}, \bibinfo{pages}{2547} (\bibinfo{year}{2009}).

\bibitem[{\citenamefont{Steele et~al.}(2009)\citenamefont{Steele, Hüttel,
  Witkamp, Poot, Meerwaldt, Kouwenhoven, and van~der Zant}}]{steele09}
\bibinfo{author}{\bibfnamefont{G.~A.} \bibnamefont{Steele}},
  \bibinfo{author}{\bibfnamefont{A.~K.} \bibnamefont{Hüttel}},
  \bibinfo{author}{\bibfnamefont{B.}~\bibnamefont{Witkamp}},
  \bibinfo{author}{\bibfnamefont{M.}~\bibnamefont{Poot}},
  \bibinfo{author}{\bibfnamefont{H.~B.} \bibnamefont{Meerwaldt}},
  \bibinfo{author}{\bibfnamefont{L.~P.} \bibnamefont{Kouwenhoven}},
  \bibnamefont{and} \bibinfo{author}{\bibfnamefont{H.~S.~J.}
  \bibnamefont{van~der Zant}}, \bibinfo{journal}{Science}
  \textbf{\bibinfo{volume}{325}}, \bibinfo{pages}{1103} (\bibinfo{year}{2009}).

\bibitem[{\citenamefont{Bouchiat et~al.}(1998)\citenamefont{Bouchiat, Vion,
  Joyez, Esteve, and Devoret}}]{bouchiat98}
\bibinfo{author}{\bibfnamefont{V.}~\bibnamefont{Bouchiat}},
  \bibinfo{author}{\bibfnamefont{D.}~\bibnamefont{Vion}},
  \bibinfo{author}{\bibfnamefont{P.}~\bibnamefont{Joyez}},
  \bibinfo{author}{\bibfnamefont{D.}~\bibnamefont{Esteve}}, \bibnamefont{and}
  \bibinfo{author}{\bibfnamefont{M.~H.} \bibnamefont{Devoret}},
  \bibinfo{journal}{Phys. Scr.} \textbf{\bibinfo{volume}{T76}},
  \bibinfo{pages}{165} (\bibinfo{year}{1998}).

\bibitem[{\citenamefont{Nakamura et~al.}(1999)\citenamefont{Nakamura, Pashkin,
  and Tsai}}]{nakamura99}
\bibinfo{author}{\bibfnamefont{Y.}~\bibnamefont{Nakamura}},
  \bibinfo{author}{\bibfnamefont{{\relax Yu}.~A.} \bibnamefont{Pashkin}},
  \bibnamefont{and} \bibinfo{author}{\bibfnamefont{J.~S.} \bibnamefont{Tsai}},
  \bibinfo{journal}{Nature} \textbf{\bibinfo{volume}{398}},
  \bibinfo{pages}{786} (\bibinfo{year}{1999}).

\bibitem[{\citenamefont{Wallraff et~al.}(2005)\citenamefont{Wallraff, Schuster,
  Blais, Frunzio, Majer, Devoret, Girvin, and Schoelkopf}}]{wallraff05}
\bibinfo{author}{\bibfnamefont{A.}~\bibnamefont{Wallraff}},
  \bibinfo{author}{\bibfnamefont{D.~I.} \bibnamefont{Schuster}},
  \bibinfo{author}{\bibfnamefont{A.}~\bibnamefont{Blais}},
  \bibinfo{author}{\bibfnamefont{L.}~\bibnamefont{Frunzio}},
  \bibinfo{author}{\bibfnamefont{J.}~\bibnamefont{Majer}},
  \bibinfo{author}{\bibfnamefont{M.~H.} \bibnamefont{Devoret}},
  \bibinfo{author}{\bibfnamefont{S.~M.} \bibnamefont{Girvin}},
  \bibnamefont{and} \bibinfo{author}{\bibfnamefont{R.~J.}
  \bibnamefont{Schoelkopf}}, \bibinfo{journal}{Phys. Rev. Lett.}
  \textbf{\bibinfo{volume}{95}}, \bibinfo{pages}{060501}
  (\bibinfo{year}{2005}).

\bibitem[{\citenamefont{Armour et~al.}(2002)\citenamefont{Armour, Blencowe, and
  Schwab}}]{armour02}
\bibinfo{author}{\bibfnamefont{A.~D.} \bibnamefont{Armour}},
  \bibinfo{author}{\bibfnamefont{M.~P.} \bibnamefont{Blencowe}},
  \bibnamefont{and} \bibinfo{author}{\bibfnamefont{K.~C.}
  \bibnamefont{Schwab}}, \bibinfo{journal}{Phys. Rev. Lett.}
  \textbf{\bibinfo{volume}{88}}, \bibinfo{pages}{148301}
  (\bibinfo{year}{2002}).

\bibitem[{\citenamefont{Tian}(2005)}]{tian05}
\bibinfo{author}{\bibfnamefont{L.}~\bibnamefont{Tian}}, \bibinfo{journal}{Phys.
  Rev. B} \textbf{\bibinfo{volume}{72}}, \bibinfo{pages}{195411}
  (\bibinfo{year}{2005}).

\bibitem[{\citenamefont{Hofheinz et~al.}(2009)\citenamefont{Hofheinz, Wang,
  Ansmann, Bialczak, Lucero, Neeley, O'Connell, Sank, Wenner, Martinis
  et~al.}}]{hofheinz09}
\bibinfo{author}{\bibfnamefont{M.}~\bibnamefont{Hofheinz}},
  \bibinfo{author}{\bibfnamefont{H.}~\bibnamefont{Wang}},
  \bibinfo{author}{\bibfnamefont{M.}~\bibnamefont{Ansmann}},
  \bibinfo{author}{\bibfnamefont{R.~C.} \bibnamefont{Bialczak}},
  \bibinfo{author}{\bibfnamefont{E.}~\bibnamefont{Lucero}},
  \bibinfo{author}{\bibfnamefont{M.}~\bibnamefont{Neeley}},
  \bibinfo{author}{\bibfnamefont{A.~D.} \bibnamefont{O'Connell}},
  \bibinfo{author}{\bibfnamefont{D.}~\bibnamefont{Sank}},
  \bibinfo{author}{\bibfnamefont{J.}~\bibnamefont{Wenner}},
  \bibinfo{author}{\bibfnamefont{J.~M.} \bibnamefont{Martinis}},
  \bibnamefont{et~al.}, \bibinfo{journal}{Nature}
  \textbf{\bibinfo{volume}{459}}, \bibinfo{pages}{546} (\bibinfo{year}{2009}).

\bibitem[{\citenamefont{Clerk}(2004)}]{clerk04}
\bibinfo{author}{\bibfnamefont{A.~A.} \bibnamefont{Clerk}},
  \bibinfo{journal}{Phys. Rev. B} \textbf{\bibinfo{volume}{70}},
  \bibinfo{pages}{245306} (\bibinfo{year}{2004}).

\bibitem[{\citenamefont{Rigas et~al.}(2006)\citenamefont{Rigas, G\"{u}hne, and
  L\"{u}tkenhaus}}]{rigas06}
\bibinfo{author}{\bibfnamefont{J.}~\bibnamefont{Rigas}},
  \bibinfo{author}{\bibfnamefont{O.}~\bibnamefont{G\"{u}hne}},
  \bibnamefont{and}
  \bibinfo{author}{\bibfnamefont{N.}~\bibnamefont{L\"{u}tkenhaus}},
  \bibinfo{journal}{Phys. Rev. A} \textbf{\bibinfo{volume}{73}},
  \bibinfo{pages}{012341} (\bibinfo{year}{2006}).

\bibitem[{\citenamefont{Flowers-Jacobs
  et~al.}(2007)\citenamefont{Flowers-Jacobs, Schmidt, and
  Lehnert}}]{flowers-jacobs07}
\bibinfo{author}{\bibfnamefont{N.~E.} \bibnamefont{Flowers-Jacobs}},
  \bibinfo{author}{\bibfnamefont{D.~R.} \bibnamefont{Schmidt}},
  \bibnamefont{and} \bibinfo{author}{\bibfnamefont{K.~W.}
  \bibnamefont{Lehnert}}, \bibinfo{journal}{Phys. Rev. Lett.}
  \textbf{\bibinfo{volume}{98}}, \bibinfo{pages}{096804}
  (\bibinfo{year}{2007}).

\bibitem[{\citenamefont{Poggio et~al.}(2008)\citenamefont{Poggio, Jura, Degen,
  Topinka, Mamin, Goldhaber-Gordon, and Rugar}}]{poggio08}
\bibinfo{author}{\bibfnamefont{M.}~\bibnamefont{Poggio}},
  \bibinfo{author}{\bibfnamefont{M.~P.} \bibnamefont{Jura}},
  \bibinfo{author}{\bibfnamefont{C.~L.} \bibnamefont{Degen}},
  \bibinfo{author}{\bibfnamefont{M.~A.} \bibnamefont{Topinka}},
  \bibinfo{author}{\bibfnamefont{H.~J.} \bibnamefont{Mamin}},
  \bibinfo{author}{\bibfnamefont{D.}~\bibnamefont{Goldhaber-Gordon}},
  \bibnamefont{and} \bibinfo{author}{\bibfnamefont{D.}~\bibnamefont{Rugar}},
  \bibinfo{journal}{Nature Physics} \textbf{\bibinfo{volume}{4}},
  \bibinfo{pages}{635} (\bibinfo{year}{2008}).

\bibitem[{\citenamefont{Doiron et~al.}(2008)\citenamefont{Doiron, Trauzettel,
  and Bruder}}]{doiron08}
\bibinfo{author}{\bibfnamefont{C.~B.} \bibnamefont{Doiron}},
  \bibinfo{author}{\bibfnamefont{B.}~\bibnamefont{Trauzettel}},
  \bibnamefont{and} \bibinfo{author}{\bibfnamefont{C.}~\bibnamefont{Bruder}},
  \bibinfo{journal}{Phys. Rev. Lett.} \textbf{\bibinfo{volume}{100}},
  \bibinfo{pages}{027202} (\bibinfo{year}{2008}).

\bibitem[{\citenamefont{Korotkov and Averin}(2001)}]{korotkov01_2}
\bibinfo{author}{\bibfnamefont{A.~N.} \bibnamefont{Korotkov}} \bibnamefont{and}
  \bibinfo{author}{\bibfnamefont{D.~V.} \bibnamefont{Averin}},
  \bibinfo{journal}{Phys. Rev. B} \textbf{\bibinfo{volume}{64}},
  \bibinfo{pages}{165310} (\bibinfo{year}{2001}).

\bibitem[{\citenamefont{MacDonald}(1949)}]{macdonald49}
\bibinfo{author}{\bibfnamefont{D.~K.~C.} \bibnamefont{MacDonald}},
  \bibinfo{journal}{Rep. Prog. Phys.} \textbf{\bibinfo{volume}{12}},
  \bibinfo{pages}{56} (\bibinfo{year}{1949}).

\bibitem[{\citenamefont{Schoelkopf et~al.}(1997)\citenamefont{Schoelkopf,
  Burke, Kozhevnikov, Prober, and Rooks}}]{schoelkopf97}
\bibinfo{author}{\bibfnamefont{R.~J.} \bibnamefont{Schoelkopf}},
  \bibinfo{author}{\bibfnamefont{P.~J.} \bibnamefont{Burke}},
  \bibinfo{author}{\bibfnamefont{A.~A.} \bibnamefont{Kozhevnikov}},
  \bibinfo{author}{\bibfnamefont{D.~E.} \bibnamefont{Prober}},
  \bibnamefont{and} \bibinfo{author}{\bibfnamefont{M.~J.} \bibnamefont{Rooks}},
  \bibinfo{journal}{Phys. Rev. Lett.} \textbf{\bibinfo{volume}{78}},
  \bibinfo{pages}{3370} (\bibinfo{year}{1997}).

\bibitem[{\citenamefont{\relax{Ya}. M.~Blanter and
  B\"uttiker}(2000)}]{blanter00}
\bibinfo{author}{\bibnamefont{\relax{Ya}. M.~Blanter}} \bibnamefont{and}
  \bibinfo{author}{\bibfnamefont{M.}~\bibnamefont{B\"uttiker}},
  \bibinfo{journal}{Phys. Rep.} \textbf{\bibinfo{volume}{336}},
  \bibinfo{pages}{1} (\bibinfo{year}{2000}).

\end{thebibliography}
\end{document}